\documentclass[aps,twocolumn]{revtex4-1}
\usepackage[utf8]{inputenc}
\usepackage[T1]{fontenc}
\usepackage[english]{babel}
\usepackage{amssymb,amsmath,mathtools} % Math
\usepackage{graphicx} 
\usepackage[table]{xcolor}
\usepackage{bbold}
\usepackage{braket}
\usepackage[headings]{fullpage}
\usepackage{hyperref}
\usepackage{slashed}
\usepackage[titletoc,toc,title]{appendix}
\usepackage[hang,small,bf]{caption}
\usepackage[nottoc,numbib]{tocbibind}
\usepackage[export]{adjustbox}
\usepackage{xcolor}
\usepackage{array}
\usepackage{booktabs}
\usepackage{hyperref}
\usepackage{amsthm}
\usepackage{lineno}
\usepackage{rotating}

\newcommand{\re}{{\rm Re}\,}

\newcommand{\change}[1]{\color{red}#1\color{black}}

\begin{document}
	\newcommand{\be}{\begin{eqnarray}}
		\newcommand{\ee}{\end{eqnarray}}
	\newcommand{\del}{\partial}
	\newcommand{\nn}{\nonumber}
	\newcommand{\STr}{{\rm Str}}
	\newcommand{\Sdet}{{\rm Sdet}}
	\newcommand{\Pf}{{\rm Pf}}
	\newcommand{\mat}{\left ( \begin{array}{cc}}
		\newcommand{\emat}{\end{array} \right )}
	\newcommand{\vect}{\left ( \begin{array}{c}}
		\newcommand{\evect}{\end{array} \right )}
	\newcommand{\tr}{{\rm Tr}}
	\newcommand{\hm}{\hat m}
	\newcommand{\ha}{\hat a}
	\newcommand{\hz}{\hat z}
	\newcommand{\hze}{\hat \zeta}
	\newcommand{\hx}{\hat x}
	\newcommand{\hy}{\hat y}
	\newcommand{\tm}{\tilde{m}}
	\newcommand{\ta}{\tilde{a}}
	\newcommand{\U}{\rm U}
	\newcommand{\D}{\slashed{D}}
	\newcommand{\hc}{^\dagger}
	\newcommand{\inv}{^{-1}}
	\newcommand{\diag}{{\rm diag}}
	\newcommand{\sign}{{\rm sign}}
	\newcommand{\ct}{\tilde{c}}
	\newtheorem{theorem}{Conjecture}
	\newcommand{\eins}{\text{Id}_n}
	
	\title[Duration of Transients in Outbreaks]{Duration of Transients in Outbreaks: When can Infectiousness be Estimated?}
	
	\author{Adam Mielke}\email{admi@dtu.dk}\affiliation{Department of Applied Mathematics and Computer Science, Technical University of Denmark, Asmussens Allé, 303B, 2800 Kgs.\ Lyngby, Denmark}
	\author{Lasse Engbo Christiansen}\email{lsec@ssi.dk}\affiliation{Statens Serum Institut, Artillerivej 5, 2300 Copenhagen S, Denmark}
	
	\date{\today}
	
	\begin{abstract}
		We investigate sub-leading orders of the classic SEIR-model using contact matrices from modeling of the Omicron and Delta variants of COVID-19 in Denmark. The goal of this is to illustrate when the growth rate, and by extension the infection transmission potential (basic or initial reproduction number), can be estimated in a new outbreak, e.g. after introduction of a new variant of a virus. In particular, we look at the time scale on which this happens in a realistic outbreak to guide future data collection.
		We find that as long as susceptible depletion is a minor effect, the transients are gone within around 3 weeks corresponding to about 4-5 times the incubation time. We also argue that this result generalizes to other airborne diseases in a fully mixed population.
	\end{abstract}
	
	\date{\today}
	\maketitle
	
	\section{Introduction}	
	An exponential fit is commonly used to estimate the infectiousness (e.g.\ the initial reproduction number) at the introduction of a new disease or variant. This is of course highly dependent on the growth actually being exponential in nature, and any non-exponential part of the growth therefore has significant consequences for these estimates. The assumption of exponential growth typically comes from a fully mixed model such as SIR-type models, where the exponential growth regime is well-described by the expansion around the disease-free equilibrium \cite{SIR1, SIR2, SIR3}.
	For a stratified model that considers several groups, the maximum growth rate is equal to the largest eigenvalue of the generator matrix. However, this requires the system to be in the corresponding eigenstate, which is rarely the case initially, as the introduction of a disease often comes from a few single individuals.
	It is therefore essential to understand when the transients of an outbreak have disappeared and the solution (temporarily) has converged sufficiently to the dominant eigenstate. A similar analysis can be found in \cite{RateOfConvergence}, but we here use data to quantify the rates, and also go into more details around the structure of the convergence.
    
    This is of particular interest when comparing the relative infectiousness of an emerging new variant versus existing variants, see for instance \cite{Household}. This is not to say that the competition between variants behaves the same way as mixing in age groups, but that the transients of a new variant have to disappear before the growth rates can be compared. Such an analysis is only relevant when the new variant enters before susceptible depletion affects the old one, so we can assume that both variants have plenty of susceptibles available. That is, we look at the case where the distribution of susceptibles is very close to the population distribution, such that the growth rates of two different strains can be compared. (Or we at least need to be able to reasonably compensate for the amount of infected with the old strain.) Note that, on a mathematical level our analysis only requires a relative unchanged number of susceptibles on the time scale we consider, so in model-based studies, we simply require that there is little cross-immunity, i.e., if we look at the introduction of Strain B, we require that having had the previous Strain A does not make you immune to Strain B. (Otherwise the low number of susceptibles will make the time-dependence of the generator matrix important.)
	
	Other mechanisms that result in non-exponential growth of epidemics have been discussed in the literature, in particular sub-exponential growth, both in data \cite{EarlyEpiSubExp1} and in ODE-models of the form \cite{EarlyEpiSubExp2, EarlyEpiSubExp3}
	\begin{eqnarray}\label{Eq:SubExp1}
		\dot{N}(t) &=& r N(t)^p
	\end{eqnarray}
	for $0<p<1$ or an exponentially decaying contact rate $\beta(t) = \beta_0 e^{-\kappa t}$ in an SIR-type model \cite{EarlyEpiSubExp4}. Instead, this paper focuses on the apparent lack of exponential growth during the transient part of the dynamics that are a consequence of stratification, even in a fully mixed population.
	
	We will be working in a stratified SEIR-model framework, which has been the basis of multiple efforts to predict the real-world spread of disease, especially in recent years \cite{SpatialEpidemicsPoland, SpatialEpidemicsStatMech, EkspertgruppenRapporter, SpatialEpidemicsGraph, SpatialEpidemicsFluid1, SpatialEpidemicsFluid2, SpatialEpidemicsFluid3, Liu, SIR4, SIR5}.
	That is, we start with the following system of differential equations
	\begin{align}\label{Eq:SEIR-model}
		\begin{split}
			\dot{S} &= - \diag(S)B I\\
			\dot{E} &=   \diag(S) B I - \eta E\\
			\dot{I} &= \eta E - \gamma I\\
			\dot{R} &= \gamma I
		\end{split}
	\end{align}
	The state vectors $S$, $E$, $I$, and $R$ represent the fraction of the population that are susceptible, exposed, infectious, and recovered respectively. That is, the states are normalized to
	\begin{align}
		\begin{split}
			S+E+I+R &= \nu\\
			\sum_{j=1}^{n} \nu_j &= 1
		\end{split}
	\end{align}
	where $\nu$ is the population distribution and $n$ is the number of groups.
	The scalar parameters $\eta$ and $\gamma$ are rates for the incubation and infection states respectively. (I.e., for simplicity, we assume these to be the same for all groups.) The matrix $B$ contains the contacts between each group. These groups can be for instance be age groups, species, or location, age groups being the most relevant of which. We do not consider transitions between groups, because we are interested in a time scale, where these are negligible. (E.g., age groups are on the order of years, whereas the period of time where exponential growth is a reasonable approximation for an acute infection like COVID-19 is on the order of weeks.) The notation $\diag(S)$ indicates a diagonal matrix with the vector $S$ on the diagonal.
	
	We will start with some standard theoretical observations about time scales. Then we will argue that the generality of these considerations extends beyond COVID-19, though the COVID-19 pandemic, specifically in Denmark, is where we make estimates. Finally, we will illustrate the convergence with numerical solution of Equation \eqref{Eq:SEIR-model} while comparing to the theoretical time scale.
	
	\section{Time Scale of Transients}
	We start by making a standard linearization by assuming that $S(t)$ changes slowly \cite{SIR1, SIR2, SIR3}. This is also the starting point of exponential growth.
	This means we have a generator matrix of the form
	\begin{eqnarray}\label{Eq:SEIR:GenMat}
		G &=& \left(\begin{matrix}
			-\eta\ \eins & \diag(S)B\\
			\eta\ \eins & -\gamma\ \eins
		\end{matrix}\right)
	\end{eqnarray}
	for the column vector $\phi(t) = \left(\begin{matrix} E(t)\\ I(t) \end{matrix}\right)$, where $\eins$ is the $n$-dimensional identity matrix. That is,
	\begin{eqnarray}
		\dot{\phi}(t) &=& G \phi(t)
	\end{eqnarray}
	Given some initial condition $\phi(0)$, we make the decomposition
	\begin{eqnarray}
		\phi(t) &=& \sum_j \alpha_j v_j e^{r_j t}
	\end{eqnarray}
	for small $t$, where $v_j$ is the $j'th$ eigenvector of $G$ and $r_j$ is the corresponding eigenvalue.
	The coefficients are determined by the initial condition and the contact matrix
	\begin{eqnarray}
		\alpha_j = \left[\left(\begin{matrix}
			| & | & |\\
			v_1 & \dots & v_N\\
			| & | & |
		\end{matrix}\right)\inv \phi(0)\right]_j\ .
	\end{eqnarray}
	Note that because the generator matrix is non-Hermitian (real, non-symmetric), the eigenvectors do not form an orthonormal basis, which means that the coefficients $\alpha_j$ are not simply the overlap between the corresponding eigenvector and the state $(v_j)^T\phi(0)$ as it would be for a Hermitian matrix. The observed growth rate becomes
	\begin{eqnarray}
		r_{\rm obs}\change{(t)} = \frac{d}{dt}\ln\left(\sum_k \phi_k(t)\right) = \frac{\sum_{j} \sum_{k} \alpha_j r_j (v_j)_k e^{r_j t}}{\sum_{j}\sum_{k} \alpha_j (v_j)_k e^{r_j t}}\ .\nn\\
	\end{eqnarray}
	Let us arrange the eigenvalues such that $r_1$ is the largest real eigenvalue and rewrite this as
	\begin{eqnarray}\label{Eq:r_timescale}
		r_{\rm obs}\change{(t)} &=& \frac{\alpha_1 r_1 \sum_{k} (v_1)_k + \sum_{j>1} \sum_{k} \alpha_j r_j (v_j)_k e^{(r_j - r_1) t}}{\alpha_1 \sum_{k} (v_1)_k + \sum_{j>1} \sum_{k} \alpha_j (v_j)_k e^{(r_j - r_1) t}}\ .\nn\\
	\end{eqnarray}
	Clearly, the components $r_j,\ j>1$ are exponentially suppressed for large enough separation between the eigenvalues, but as long as the time is on the same scale as the inverse of this separation, non-equilibrium effects are visible. The time scale $\tau = \frac{1}{r_1 - \max_{j>1}\re(r_j)}$ thus sets the upper bound of the convergence time, where convergence occurs at time $t\gg\tau$. Note that this is an upper bound as faster decaying contributions may be the largest ones in the beginning. That is, the balance between $\alpha_j$ and $r_j$ in Equation \eqref{Eq:r_timescale} may be such that the convergence is faster than this scale $\tau$.
	
	Note that $\tau$ is primarily determined by the contacts and susceptibles $\diag(S)B$ and incubation time $1/\eta$ rather than the infectious period $1/\gamma$. To see this, note that the eigenvalues of $G$ may be expressed in terms of the eigenvalues $x_j$ of $\diag(S) B$
	\begin{eqnarray}\label{Eq:Eigenvalues}
		r_j &=& \frac{-\eta-\gamma\pm\sqrt{(\eta-\gamma)^2 + 4\eta x_j}}{2}\ .
	\end{eqnarray}
    Each eigenvalue of $\diag(S) B$ corresponds to two eigenvalues of $G$, which is in line with $G$ being twice the size of $\diag(S) B$. 
        
    The $B$-matrix has a covariance matrix structure arising from the fact that individuals have to be at the same place at the same time. That is, if we discretize space-time into $K$ points, the matrix $C$ containing the number of contacts between groups may abstractly be expressed as
    \begin{eqnarray} 
        C &=& WW^T
    \end{eqnarray}
    where the $n\times K$ matrix $W$ is a time series that is non-zero at the space and time that people meet.
    The contact (rate) matrix $B$ is then given as
    \begin{eqnarray} \label{Eq:Bstructure}
        C &=& \diag(N)B\diag(N)\nn\\
        \Rightarrow B &=& \diag(N)\inv WW^T \diag(N)\inv 
    \end{eqnarray}
    where $N$ is a vector containing the populations of each group. (So $N$ is proportional to $\nu$.) We can absorb $\diag(N)$ in $W$ if need be.
    
	For those familiar with the Altland-Zirnbauer classification of operators \cite{Zirnbauer, AltlandZirnbauer1, AltlandZirnbauer2}, the $B$-matrix is sampled from a subset of the real Wishart-Laguerre Ensemble because of its covariance matrix structure.
	However, it is too great a simplification not to assume some substructure, so we instead use the different lockdown conditions to represent our ensemble.
    
    The eigenvalues of $\diag(S) B$ also behave like this ensemble (though not the eigenvectors) because of the similarity transformation
	\begin{eqnarray}
		&&\diag(\sqrt{S})\inv \diag(S)B \diag(\sqrt{S})\nn\\
		&=& \diag(\sqrt{S}) B \diag(\sqrt{S})\nn\\
        &=& \left(\diag(\sqrt{S}) W\right) \left(W^T\diag(\sqrt{S})\right) \ .
	\end{eqnarray}
    where we can absorb $\diag(\sqrt{S})$ in $W$. (The square root of the vector $S$ should be taken element-wise.) 
	This fixes all $x_j$ to be real and non-negative.
	Equation \eqref{Eq:Eigenvalues} comes directly from applying the block matrix property
	\begin{eqnarray}
		\det\left(\begin{matrix}
			M_1 & M_2\\
			M_3 & M_4
		\end{matrix}\right) &=& \det\left(M_1 M_4 - M_2 M_3\right)
	\end{eqnarray}
	for commuting $M_3$ and $M_4$ to the eigenvalue equation. (This requires that $\diag(S)B$ is diagonalizable, which follows from Equation \eqref{Eq:Bstructure}.)
	The time scale thus becomes
	\begin{eqnarray}\label{Eq:ScalingOfTau}
		\tau &=& \frac{1}{r_1 - r_2}\nn\\
		&=& \frac{1}{\sqrt{\eta}} \frac{1}{\sqrt{\frac{(\eta-\gamma)^2}{4\eta} + x_1} - \sqrt{\frac{(\eta-\gamma)^2}{4 \eta} + x_2}}\ .
	\end{eqnarray}
	As $\eta$ and $\gamma$ typically are of the same scale, $\frac{(\eta-\gamma)^2}{4\eta}\ll 1$, the time scale $\tau$ primarily scales with the square root of the incubation time $1/\sqrt{\eta}$ and the differences of the square roots of eigenvalues of $\diag(S)B$. (A deformed spectral gap, in a sense.)
	
	The specific relationship between $r_j$ and $x_j$ in Equation \eqref{Eq:Eigenvalues} is of course model-dependent, so let us look at what happens if more sub-states (like $S\to E_1\to E_2\to I_1\to I_2$) are added
	\begin{eqnarray}\label{Eq:SEIR:GenMat2}
		\widetilde{G} &=& \left(\begin{matrix}
			-2\eta\ \eins & 0 & \diag(S)B & \diag(S)B\\
			2\eta\ \eins  & -2\eta\ \eins & 0 & 0\\
			0 & 2\eta\ \eins & -2\gamma\ \eins & 0\\
			0 & 0 & 2\gamma\ \eins & -2\gamma\ \eins
		\end{matrix}\right)\nn\\
	\end{eqnarray}
	We can diagonalize $\diag(S)B$ directly because the remaining blocks are proportional to the identity matrix. We can additionally rearrange rows and columns to make $\widetilde{G}$ block-diagonal
	\begin{align}
		\begin{split}
			\widetilde{G} &= \left(\begin{matrix}
				\widetilde{G}_1 & &\\
				&\ddots&\\
				&&\widetilde{G}_n
			\end{matrix}\right)\ ,\\
			\widetilde{G}_j &= \left(\begin{matrix}
				-2\eta & 0 & x_j & x_j\\
				2\eta  & -2\eta & 0 & 0\\
				0 & 2\eta & -2\gamma & 0\\
				0 & 0 & 2\gamma & -2\gamma
			\end{matrix}\right)
		\end{split}
	\end{align}
	Finding the roots of the characteristic polynomial of $\widetilde{G}$ analytically does not seem feasible, but for $x_j = 0$, the eigenvalues are $-2\eta$ and $-2\gamma$, a structure shared by any number of sub-states.
	Note that $\widetilde{G}_j$ is defective for $x_j=0$, so perturbation theory cannot be applied, but continuity of the eigenvalues still suggests that the eigenvalues of $\widetilde{G}$ are in a neighborhood around $-2\eta$ and $-2\gamma$ for small $x_j$. Numerics confirm that the largest eigenvalue of $\widetilde{G}$, which we again denote by $r_j$, depends on $x_j$ to some fractional power, suggesting a relation very similar to Equation \eqref{Eq:Eigenvalues}.
	
	\section{Generality Considerations and Existence of Spectral Gap}
	It is of course important to understand whether the size of the transient scale applies to other diseases than COVID-19, which we estimate here. It is especially important, since the spectral gap is very hard to measure at the start of an outbreak.
	
	Our main argument goes as follows: We conjecture that for similar diseases (e.g.\ airborne diseases that most of the population are susceptible to), most of the structure of the contact matrix comes from societal structures. That is, the main difference between two diseases/variants is an overall factor of infectiousness, i.e., a factor on $B$, which in turn will be a factor on the basic reproduction number.
	We also assume that the contact matrix is sufficiently filled (i.e., non-sparse). In such a case, we know from the theory of random matrices \cite{Mehta, OxfordRMT} that level repulsion will be present. That is, even if the elements of the contact matrix are approximately independently distributed, the eigenvalues will be correlated and the joint distribution of eigenvalues will disappear in a larger number of cases when the eigenvalues are identical. That is, the joint distribution of eigenvalues $p(\lambda_1, \dots, \lambda_n)$ will obey
    \begin{eqnarray}
        p(\lambda_1, \dots, \lambda_n) &=& 0\text{ for any } \lambda_j = \lambda_k
    \end{eqnarray}
    There will therefore almost surely be a spectral gap, and if the structure of the matrix is preserved, the size of the eigenvalues (and thus also of the spectral gap) will simply scale with the factor in front.
	In other words, the time scale $\tau$ will simply scale with the incubation time and the inverse square root of the infectiousness, see Equation \eqref{Eq:ScalingOfTau}. Importantly, and perhaps unsurprisingly, the transients of a faster spreading disease will die out faster as well. So the cases where convergence is slow will also be the cases where there is  more time to react.
	
	Based on these observations, it becomes interesting to investigate the possible societal structures that a disease can propagate through, and by extension, what a typical $\tau$ is. We here consider two cases of infectiousness, both based on COVID-19 in Denmark. One is the Delta Variant in June of 2021, which should be considered a very low level of infectiousness (effective reproduction number close to 1). Combined with the rather high incubation time of COVID-19, this can be interpreted as an approximate upper bound on $\tau$ for other diseases. The other is the Omicron Variant in December 2021, which is an example where time was of the essence, and it was therefore important to know the growth rate quickly. (The latter example was what originally motivated  this work.)
	
	To make these estimates, we assume the whole population is susceptible and look at the $B$-matrices as an ensemble. Though a certain amount of the population had already been infected at the introduction of the Omicron Variant, we ignore this as we are mostly interested in the different infectiousness of the two variants. The ensemble approach is again in the spirit of random matrix theory. To represent the ensemble, we use the contact matrices constructed by the Danish expert group, where different societal structures under different lockdown conditions (including open society) are used individually as an ensemble of contact matrices \cite{EkspertgruppenRapporter}. This implicitly assumes that these are a representative sample of possible societal structures, but since Denmark underwent several changes to restrictions during the pandemic, we believe that this is a reasonable assumption.
	
	The sampling from an ensemble described above is also the reason why our numerical results in the following section are given by bands rather than single lines.
	
	\begin{figure*}
		\centering
		\hspace{30pt} \textbf{Omicron Infectiousness} \hspace{80pt} \textbf{Delta Infectiousness}\\
		\begin{turn}{90}\hspace{30pt} \textbf{Start in 5-11 y.o.}\end{turn}
		\includegraphics[width=0.4\linewidth]{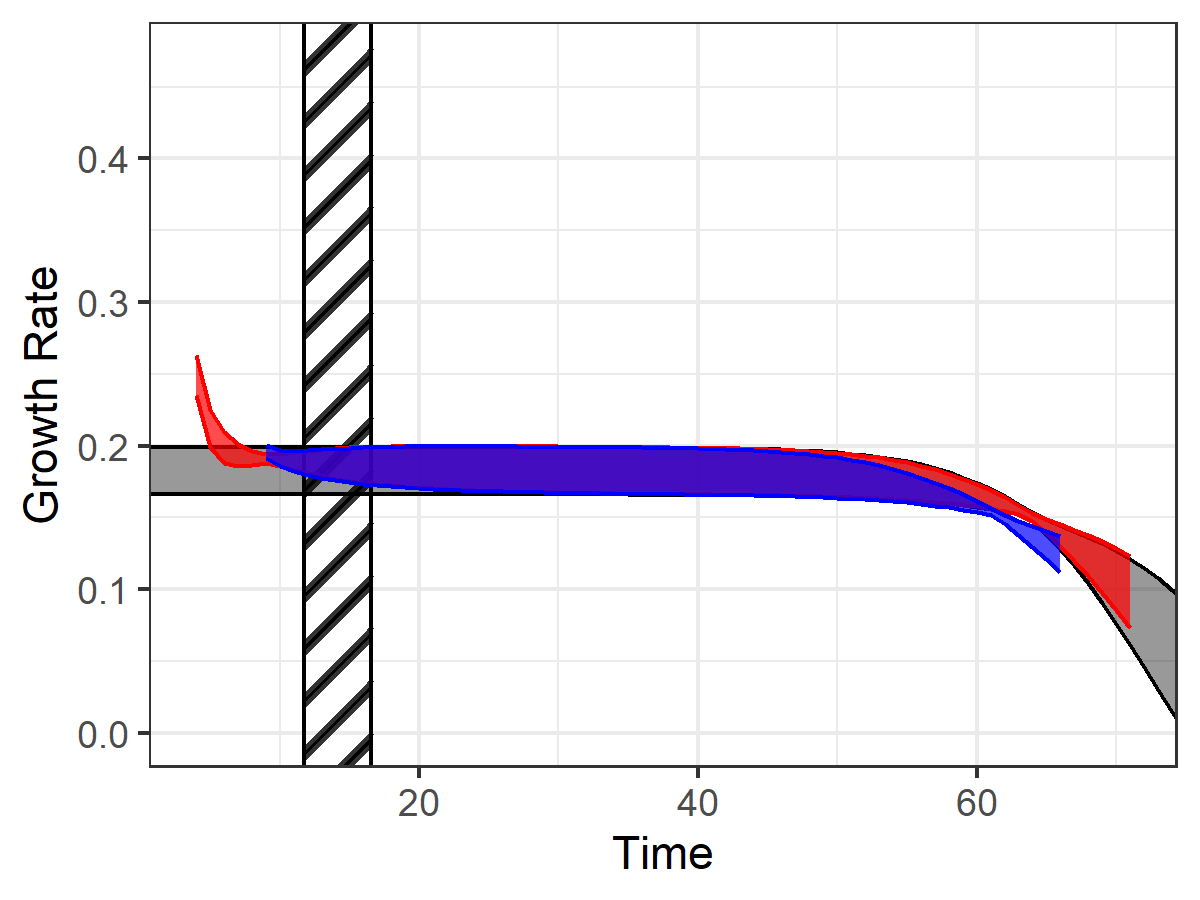}
		\includegraphics[width=0.4\linewidth]{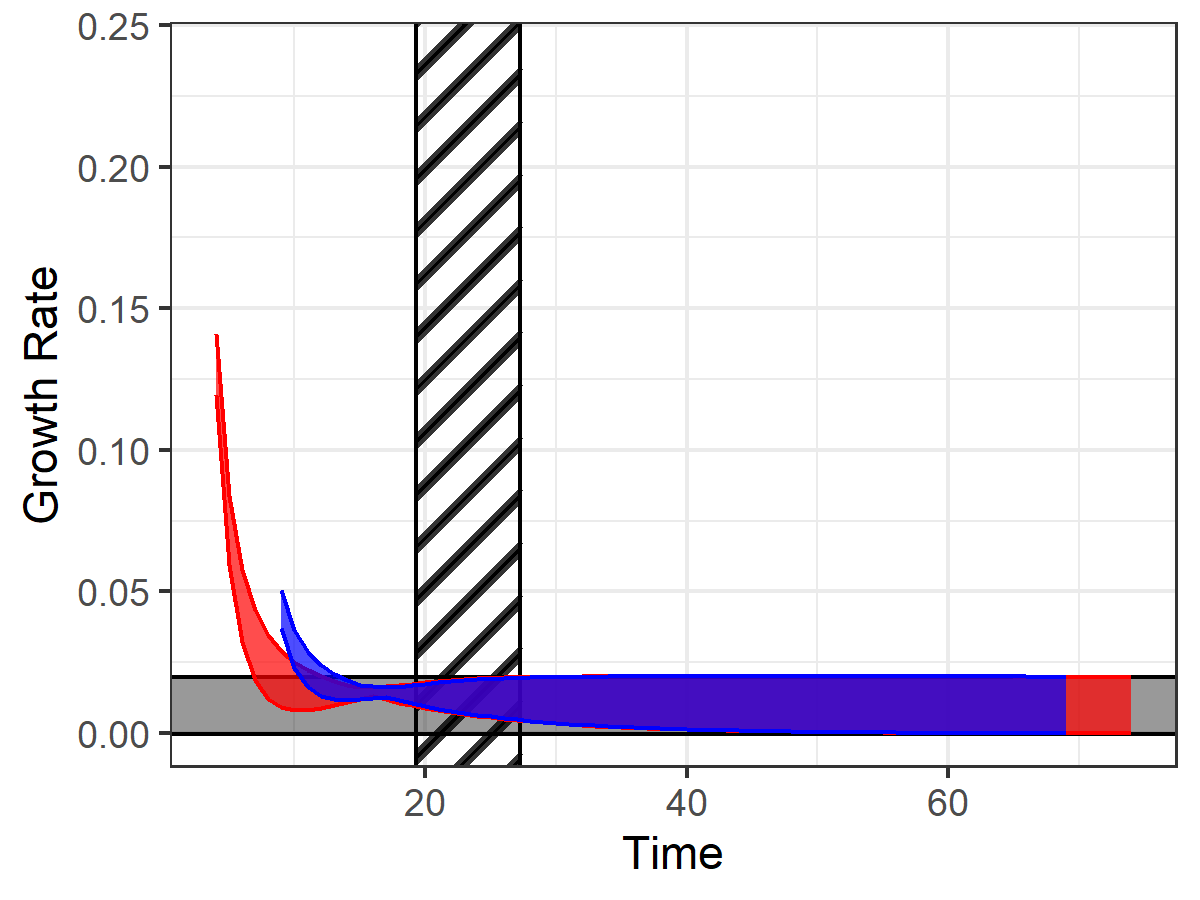}\\
		\begin{turn}{90}\hspace{30pt} \textbf{Start in 20-29 y.o.}\end{turn}
		\includegraphics[width=0.4\linewidth]{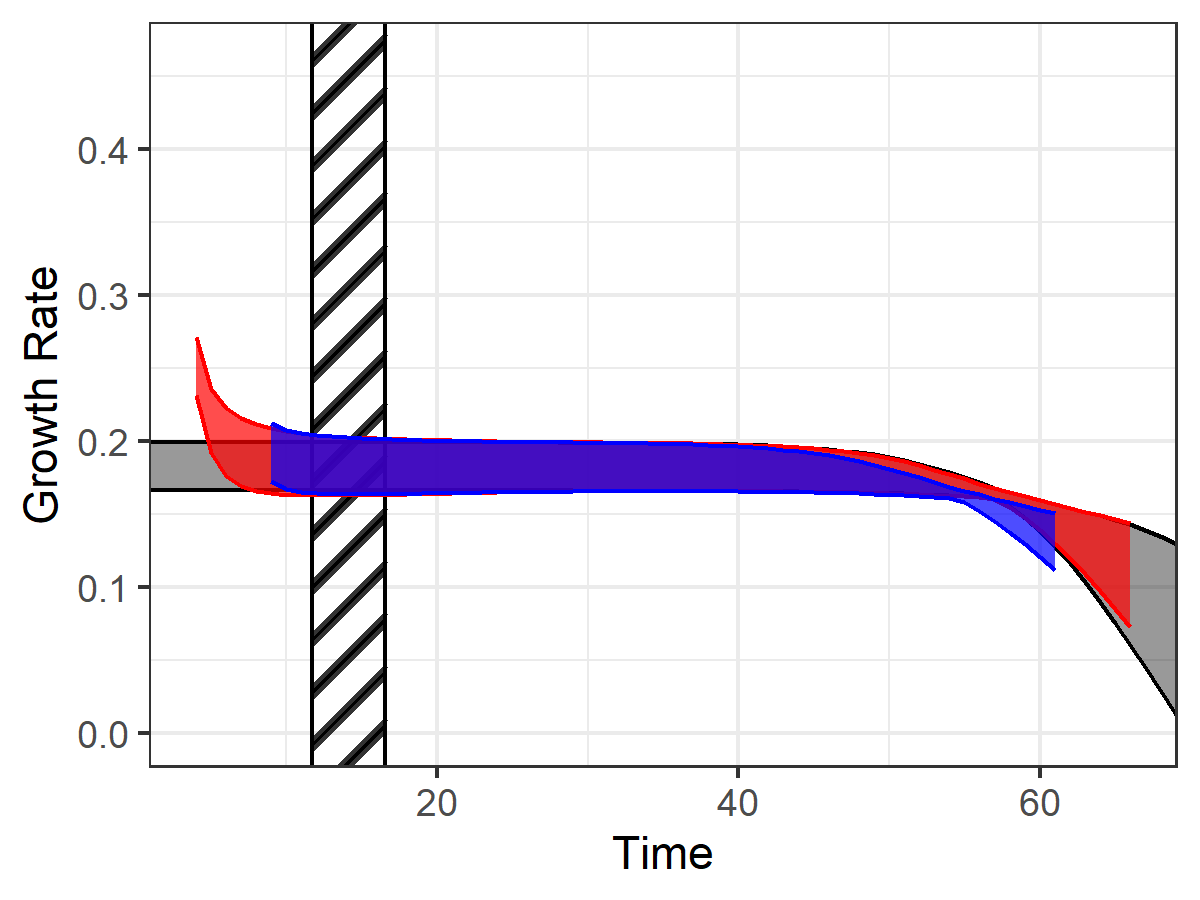}
		\includegraphics[width=0.4\linewidth]{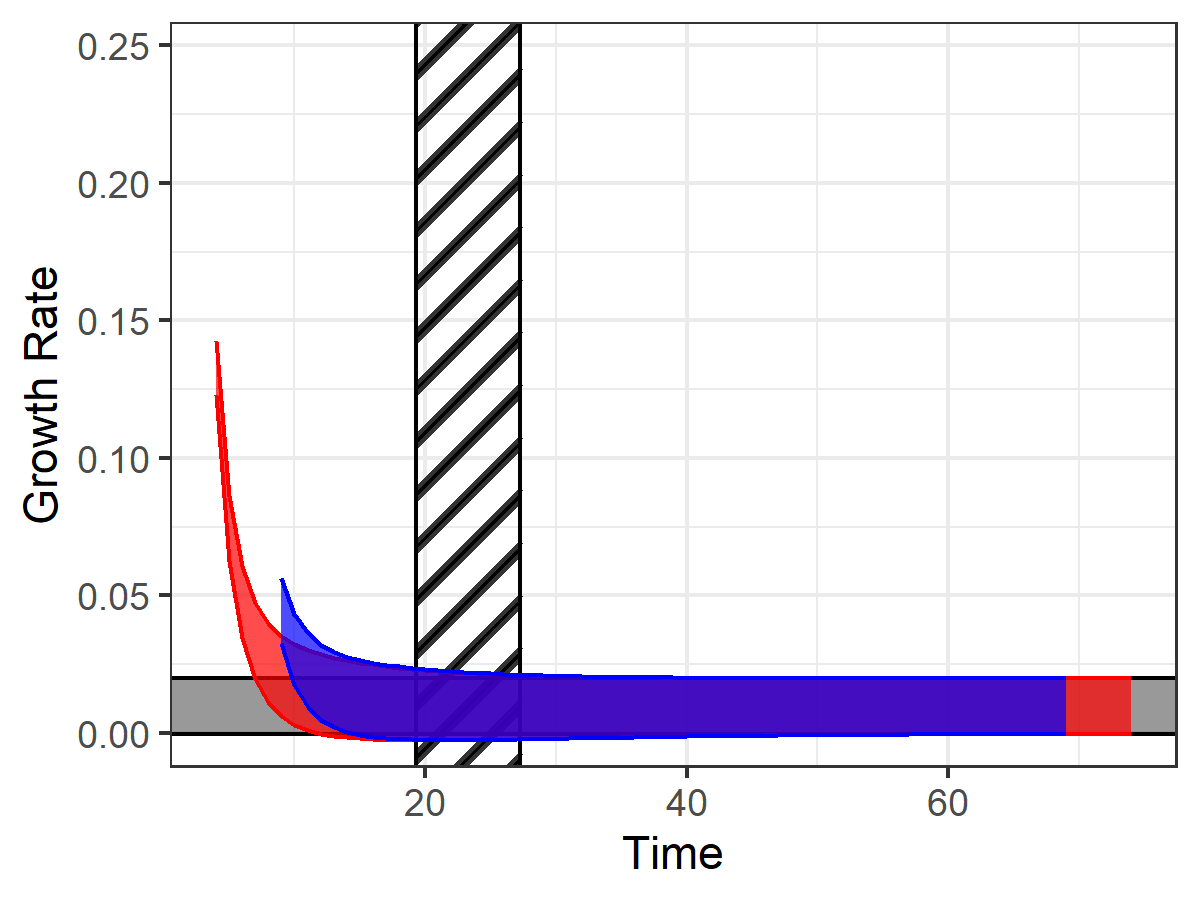}\\
		\begin{turn}{90}\hspace{30pt} \textbf{Start in 70-79 y.o.}\end{turn}
		\includegraphics[width=0.4\linewidth]{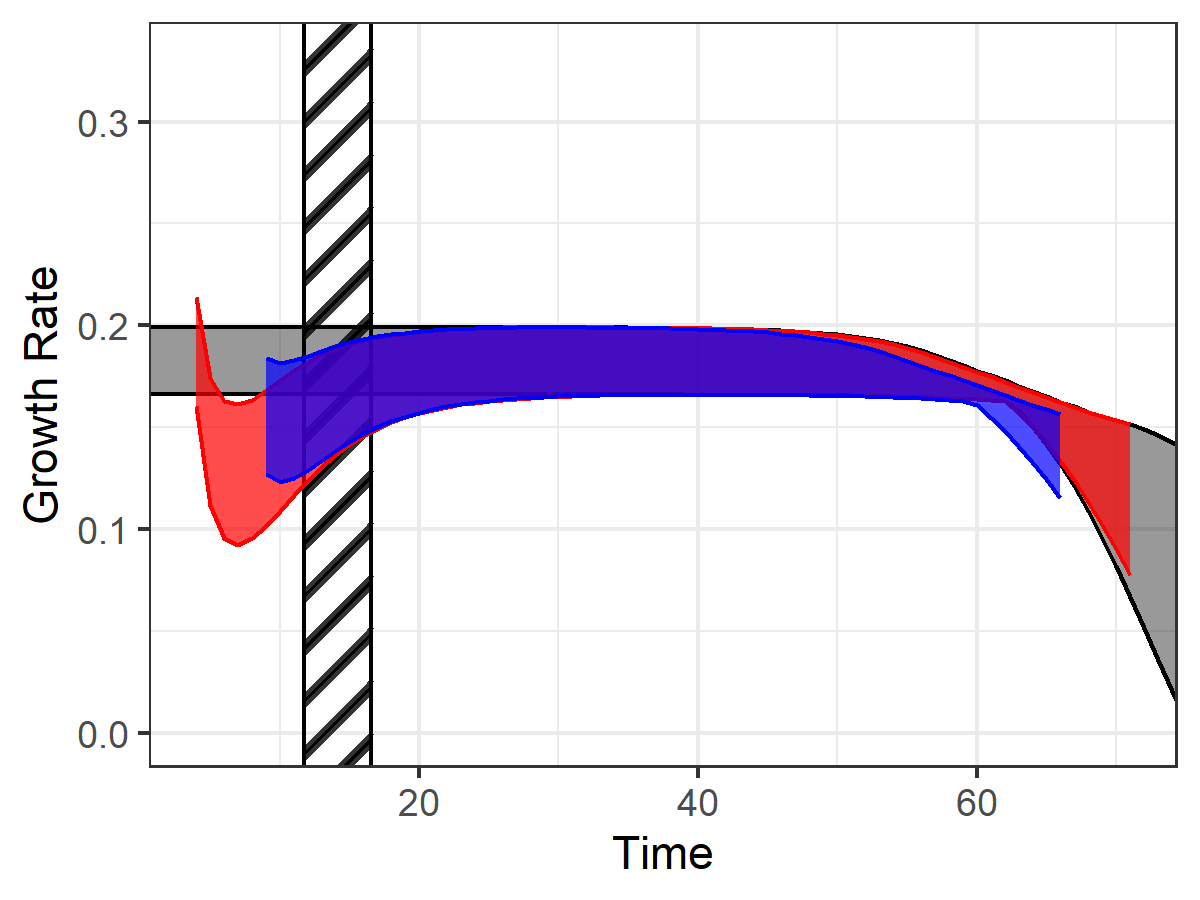}	
		\includegraphics[width=0.4\linewidth]{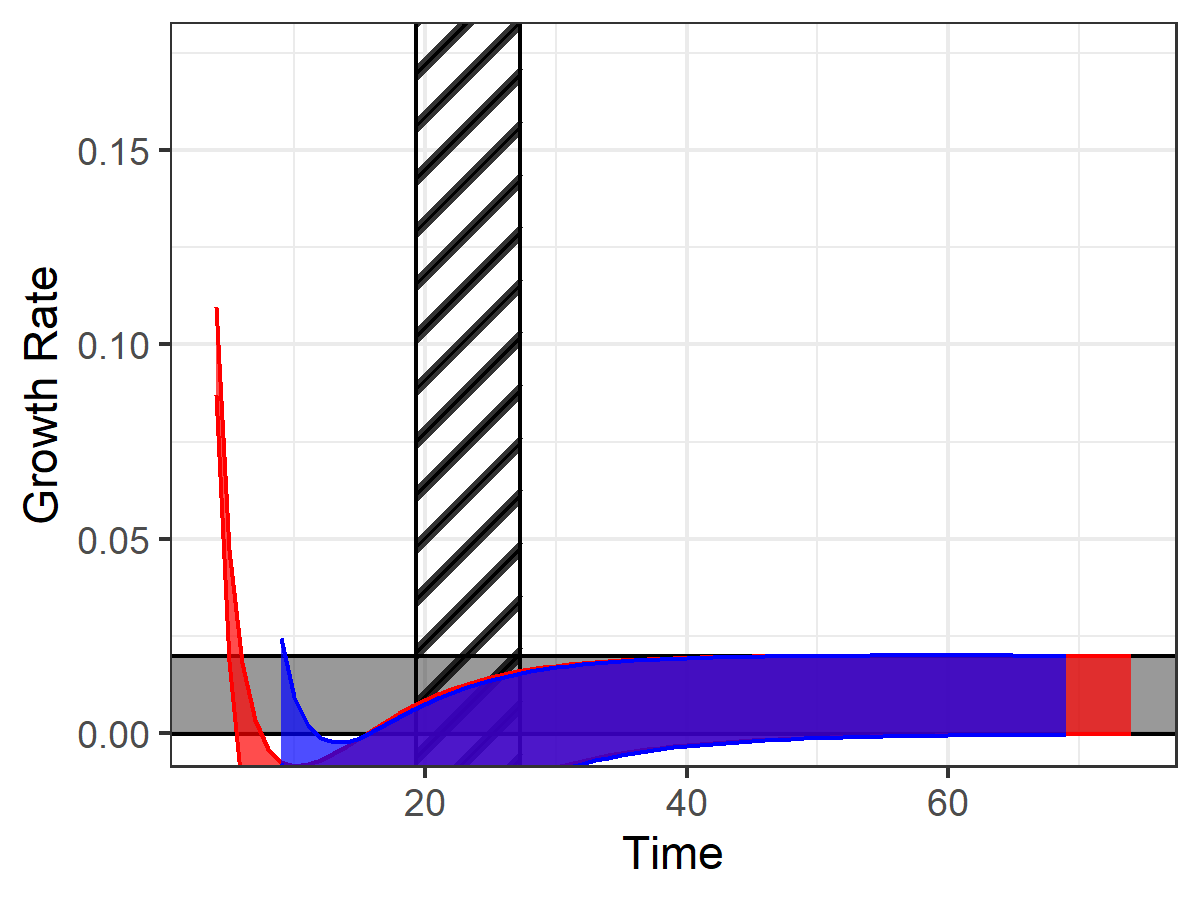}
		\caption{Estimated growth rates as a function of time for the SEIR-models given by the different contact matrices.
			The gray horizontal ribbons are the largest eigenvalue of the matrix in Equation \eqref{Eq:SEIR:GenMat}. (Note that we let $S(t)$ change with time when calculating the largest eigenvalue. Mathematically speaking, this is simply a linearization for each point in time. It therefore represent an instantaneous growth rate, which allows us to illustrate when susceptible depletion occurs.)\newline
			The red and blue ribbons are the observed growth rates, i.e.\ a linear least squares fit to $\sum_{j} I_j$ on log scale to either the 10 ($[-4, 5]$, red) or 20 ($[-9, 10]$, blue) surrounding days.
            The growth rate is associated with the middle of the interval rather than the end to align with theoretical values. Note that the growth rates of $\phi$ and $I$ are the same. The striped vertical band indicates the $10\%$-point of $\max_{j>1}\exp\left[(\re(r_j) - r_1)t\right]$ corresponding to $\ln(10)\tau$ to illustrate the time scale set by Equation \eqref{Eq:r_timescale}. Each property becomes a ribbon of minimum and maximum value rather than a line of a single value because of ensemble interpretation.\newline
			The rows are initialization of the outbreak in different age groups, and the columns are normalization of the largest eigenvalue to match the infectiousness of either the Omicron variant in December 2021 or Delta variant in June 2021 in Denmark \cite{EkspertgruppenRapporter}.}
		\label{Fig:AllGR}
	\end{figure*}
	
	\section{Numerical Illustration of Solution Convergence}
	For each matrix in the ensemble, the largest eigenvalue is normalized to the infectiousness of either the Delta or the BA.1 Omicron variant at seasonal strength equivalent to the start of December. The parameters $\eta = 1/4.3$ and $\gamma = 1/3.3$ are used, consistent with \cite{EkspertgruppenRapporter}. The population is divided into 10 age groups, namely ages 0-4, 5-11, 12-19, 20-29,... , 70-79, 80+, according to the age distribution of the population of Denmark.
	We initialize with $10^{-5}$ of the age group population in the respective $E$-state in three different age groups, one at a time, and follow the progression of the observed growth rate over time. That is, a fit on log scale to the 10 or 20 surrounding days. The younger age groups are usually more active, and we therefore expect the growth rate of outbreaks starting in those groups to converge to the dominant eigenvalue faster. This turns out to be the case, and the convergence consistently happens within the first three weeks, see Figure \ref{Fig:AllGR} for an illustration.
	
	It is clear that an outbreak starting in a less active group will take longer to find the equilibrium state than one starting in the more active group. It is also clear that a higher growth rate leads to faster convergence. Neither of these facts are surprising, but the scale that it happens on is significant. It shows that waiting three weeks before measuring the growth rate is enough to negate this non-equilibrium effect, and it is therefore safe to estimate the growth rate at this point. This effect does not depend on the infectiousness, only the spacing between the eigenvalues as illustrated in Equation \eqref{Eq:r_timescale}. Note also that the difference between 10 and 20 surrounding days is only a slight shift in the estimated growth rate.

	\section{Discussion}
	We see that 10 days fit interval gives the fastest convergence, but in reality data should be expected to have weekly variations, both in reporting testing patterns and behavior, which means that at least 14 or 21 days of data are needed to get robust estimates of the growth rates.
	Interestingly, the growth rate starts above the expected irrespective of which age group is infected first (compare the three rows of Figure \ref{Fig:AllGR}). This is not possible in the framework that Equation \eqref{Eq:r_timescale} represents, as it is a weighted average of the different eigenvalues. The larger growth rate instead comes from the differences between the finite times of measuring (i.e.\ once a day) and the instantaneous nature of an ODE-model. That is, the relative change becomes singular at the origin (not shown). This also means that smaller time steps do not help.
	The convergence to the largest eigenvalue of course also depends on the initial conditions. Interestingly, the approach from above is not only an effect of starting in the $E$-state, as this can occur without anyone in the $E$-state initially, i.e., with the entire infected population starting in the $I$-state. The specific initialization at $10^{-5}$ does not affect the results. Initializing at $10^{-6}$ makes an almost imperceptible difference to the growth rate dynamics.
	These sensitivity analyses are described qualitatively instead of plotted for the sake of brevity.
	
	In a realistic setting, some groups are more likely to be tested than others so the observed incidences may misrepresent in which group the disease is most prevalent. If there are significant delays in testing, this will of course also affect the time scale, as we here look at the point where people turn infectious.
	
	For a single point of introduction, the time scale on which convergence happens is of the same order as that of the spatial effects, compare the time scales of Figure \ref{Fig:AllGR} left column in this paper to that of Figure 1 in \cite{SecondQuant}, which is an analysis of the same period of time in Denmark. This means that both contributions have to be considered simultaneously. However, spatial dynamics and network models also include local outbreaks that have a brief, but fierce growth, even on a population level if the prevalence is low \cite{alphaSpread}.
	
	Some comments on the limitations of this analysis: It is common to introduce a spatial component by discretizing space and making each section of space a group, see for instance \cite{SpatialEpidemicsStatMech, Martcheva}. The time scale of transients considered here should not be expected to apply to such models, although the overall structure is the same. The reason for this is that such spatial models can undergo local depletion of susceptibles, which makes the dominant eigenvector a less useful quantity.
	
	Additionally, we here assume an age distribution similar to Denmark. A change to the vector $S$ would also result in a change in the spectral gap. We have here assumed that such changes are minor, and our results may easily be adapted to these differences.
	
	\section{Conclusion}
	In the above we have investigated the transient effects of SEIR-models when initializing the system away from the dominant eigenvector and the consequences for the observed growth rate.
	
	As expected, an outbreak starting in the more active groups converges faster, and the same is true for a more infectious disease. The significant part is the time scale on which the convergence occurs. It turns out that, using realistic parameters, the growth rate has converged within three weeks, corresponding to 4-5 times the incubation time, which also means that it is safe to estimate it at this point.
	
	We have also argued that, though these estimates are made on the COVID-19 pandemic in Denmark, the results are applicable to the introduction of other respiratory diseases with a similar distribution of susceptibles, i.e.\ low natural immunity in the population.

	\acknowledgements
	Funding and data provided by Statens Serum Institut, Denmark.


\begin{thebibliography}{99}		
		\bibitem{SIR1}
		R.\ Ross, \textit{An application of the theory of probabilities to the study of a priori pathometry - Part I}, Proc.\ R.\ Soc.\ Lond.\ A92204–230 (1916).
		
		\bibitem{SIR2}
		G.\ Bastin, \textit{Lectures on Mathematical Modelling of Biological Systems}, (2012).
		
		\bibitem{SIR3}
		H.\ Weiss, \textit{The SIR model and the Foundations of Public Health}, Materials Matem\`{a}tics 0 (2013).

        \bibitem{RateOfConvergence}
        C.\ A.\ Rhodes and T.\ House,
        \textit{The rate of convergence to early asymptotic behaviour in age-structured epidemic models,}
        Theoretical Population Biology 85, 58-62 (2013).
        
  
		\bibitem{Household}
		F.\ P.\ Lyngse et al.,
		\textit{Household transmission of the SARS-CoV-2 Omicron variant in Denmark},
		Nat.\ Commun.\ 13 (1), 5573 (2022).
		
		\bibitem{EarlyEpiSubExp1}
		G.\ Chowell, C.\ Viboud, J.\ M.\ Hyman, and L.\ Simonsen,
		\textit{The Western Africa ebola virus disease epidemic exhibits both global exponential and local polynomial growth rates},
		PLoS Curr.\ (2015).
		
		\bibitem{EarlyEpiSubExp2}
		C.\ Viboud, L.\ Simonsen, G.\ Chowell,
		\textit{A generalized-growth model to characterize the early ascending phase of infectious disease outbreaks},
		Epidemics 15, 27-37 (2016).
		
		\bibitem{EarlyEpiSubExp3}
		G.\ Chowell, C.\ Viboud, L.\ Simonsen, and S.\ M.\ Moghadas,
		\textit{Characterizing the reproduction number of epidemics with early subexponential growth dynamics},
		J.\ R.\ Soc.\ Interface 13, 123 (2016).
		% All look at model C'(t) = r C(t)^p to classify sub-exponential growth.
		
		\bibitem{EarlyEpiSubExp4} % Exponentially decaying contact rate (from behavioural response)
		F.\ Brauer,
		\textit{The Final Size of a Serious Epidemic},
		Bull.\ Math.\ Biol.\ 81, 869–877 (2019). 
		
		\bibitem{Liu}
		Q.-H.\ Liu, M.\ Ajelli, A.\ Aleta, S.\ Merler, Y.\ Moreno, and A.\ Vespignani, \textit{Measurability of the epidemic reproduction number in data-driven contact networks}, PNAS 115(50)12680-12685, (2018).
		
		\bibitem{SpatialEpidemicsGraph}
		A.\ Nava, A.\ Papa, M.\ Rossi, and D.\ Giuliano, \textit{Analytical and cellular automaton approach to a generalized SEIR model for infection spread in an open crowded space} Phys.\ Rev.\ Research 2, 043379 (2020).
		
		\bibitem{SpatialEpidemicsFluid1}
		G.\ Bertaglia and L.\ Pareschi, \textit{Hyperbolic models for the spread of epidemics on networks: kinetic description and numerical methods}, ESAIM: M2AN 55, 2, 381-407 (2021).
		\bibitem{SpatialEpidemicsFluid2}
		Q.\ Zhuang and J.\ Wang, \textit{A spatial epidemic model with a moving boundary Infectious Disease Modelling}, 6, 1046-1060 (2021).
		
		\bibitem{SpatialEpidemicsStatMech}
		Ó.\ Toledano, B.\ Mula, S.\ N.\ Santalla, Javier Rodríguez-Laguna, and Ó.\ Gálvez, \textit{Effects of confinement and vaccination on an epidemic outburst: a statistical mechanics approach},
		Phys.\ Rev.\ E 104, 034310 (2021).
		
		\bibitem{SpatialEpidemicsFluid3}
		A.\ Triska, A.\ Y.\ Gunawan, N.\ Nuraini,
		\textit{Outbreak spatial pattern formation based on an SI model with the infected cross-diffusion term}, J. Math. Computer Sci., 27 1–17 (2022).
		
		\bibitem{SpatialEpidemicsPoland}
		P.\ A.\ Werner, M.\ Ksik-Brodacka, K.\ Nowak, R.\ Olszewski, M.\ Kaleta, and D.\ T.\ Liebers, \textit{Modeling the Spatial and Temporal Spread of COVID-19 in Poland Based on a Spatial Interaction Model},	ISPRS Int.\ J.\ Geo-Inf., 11, 195 (2022).
		
		\bibitem{SIR4}
		M.\ Cabrera, F.\ Córdova-Lepe, J.\ P.\ Gutiérrez-Jara et al., \textit{An SIR-type epidemiological model that integrates social distancing as a dynamic law based on point prevalence and socio-behavioral factors}, Sci Rep 11, 10170 (2021).
		
		\bibitem{SIR5}
		M.\ Bisiacco and G.\ Pillonetto, \textit{COVID-19 epidemic control using short-term lockdowns for collective gain},	Annual Reviews in Control, 52, 573-586 (2021).
		
		\bibitem{EkspertgruppenRapporter}
		https://covid19.ssi.dk/analyser-og-prognoser/modelberegninger (In Danish, visited 2022-12-21).
		
		\bibitem{Zirnbauer}
		M.\ R.\ Zirnbauer,
		\textit{Riemannian Symmetric Superspaces and their Origin in Random Matrix Theory},
		J.\ Math. Phys.\ 37, 4986 (1996).
		
		\bibitem{AltlandZirnbauer1}
		A.\ Altland and M.\ R.\ Zirnbauer,
		\textit{Random Matrix Theory of a Chaotic Andreev Quantum Dot},
		Phys. Rev. Lett. 76, 3420 (1996).
		
		\bibitem{AltlandZirnbauer2}
		\textit{Novel Symmetry Classes in Mesoscopic Normal-Superconducting Hybrid Structures},
		Phys.\ Rev.\ B 55, 1142 (1997).
		
		\bibitem{Mehta} M.\ L.\ Mehta, \textit{Random Matrices}, (Third Edition, Elsevier, 2004).
		
		\bibitem{OxfordRMT} G. Akemann, J. Baik, and P. Di Francesco (eds.),
		\textit{The Oxford Handbook of Random Matrix Theory}
		(First Edition,	Oxford University Press, 2011).
		
		\bibitem{SecondQuant}
		A.\ Mielke,
		\textit{On the Role of Spatial Effects in Early Estimates of Disease Infectiousness: A Second Quantization Approach},
		arXiv:2205.15718 [q-bio.PE] (2022), Submitted for publication.
		
		\bibitem{alphaSpread}
		T.\ Y.\ Michaelsen et al.,
		\textit{Introduction and transmission of SARS-CoV-2 lineage B.1.1.7, Alpha variant, in Denmark},
		Genome Medicine 14 (47) (2022).
		
		\bibitem{Martcheva}
		M.\ Martcheva, \textit{An Introduction to Mathematical Epidemiology}, (First Edition, Springer Science+Business Media New York, 2015).
	\end{thebibliography}
\end{document}